\documentclass[mathleft
]{an}
\usepackage{graphicx}
\usepackage{times}
\begin{document}

\Pagespan{249}{}
\Yearpublication{2009}%
\Yearsubmission{2008}%
\Month{2}%
\Volume{330}%
\Issue{2/3}%

\title{H$_2$O maser and a plasma obscuring torus in the radio galaxy NGC 1052}

\author{S. Sawada-Satoh\inst{1}\fnmsep\thanks{Corresponding author:
  \email{sss@yamaguchi-u.ac.jp}\newline},
S. Kameno\inst{2},  K. Nakamura \inst{2}, 
 D. Namikawa \inst{2},   K. M. Shibata \inst{3},  
\and  
 M. Inoue \inst{3}
 }
\titlerunning{A Dense Plasma in NGC 1052}
\authorrunning{S. Sawada-Satoh et al.}
\institute{
Department of Physics, Faculty of Science, Yamaguchi University, 
1677-1 Yoshida, Yamaguchi, 753-8512 Japan
\and 
Department of Physics, Faculty of Science, Kagoshima University, 
1-21-35 Korimoto, Kagoshima, 890-0065, Japan
\and 
National Astronomical Observatory, Mitaka, Tokyo 181-8588, Japan}

\received{8 Dec 2008}
\accepted{18 Dec 2008}
\publonline{15 Feb 2009}


\abstract{%
We present multi-frequency simultaneous VLBA observations 
at 15, 22 and 43 GHz 
towards the nucleus of the nearby radio galaxy NGC~1052.  
These three continuum images reveal a double-sided jet structure, 
whose relative intensity ratios imply 
that the jet axis is oriented close to the sky plane. 
The steeply rising spectra at 15--43 GHz at the inner edges of the jets 
strongly suggest that synchrotron emission 
is absorbed by foreground thermal plasma.  
We detected H$_2$O maser emission in the velocity range of 1550--1850 
km~s$^{-1}$, which is redshifted by 50--350 km~s$^{-1}$ 
with respect to the systemic velocity of NGC~1052. 
The redshifted maser gas appears projected against both sides of the jet, 
similar to the HI seen in absorption. 
The H$_2$O maser gas is located where the free-free absorption opacity is large. 
This probably implies that the masers in NGC~1052 are associated with 
a circumnuclear torus or disk as in the nucleus of NGC~4258. 
Such circumnuclear structure can be the sense of accreting onto the central engine.
}

\maketitle

\section{Introduction}

The elliptical galaxy NGC 1052 with a LINER is the nearest object among radio-loud and broad megamaser sources  (e.g. Gabel et al. 2000). 
The redshift of $z=0.0049$ (Knapp et al. 1978) corresponds to a distance of 20.3 Mpc and the angular scale is 98 pc arcsec$^{-1}$ if we adopt the 
Hubble constant $H_{0}=72$~km~s$^{-1}$~Mpc$^{-1}$ and $q_0=0.5$.  (Spergel et al. 2003). 
This galaxy hosts a well defined double-sided radio jet 
of a few pc with P.A. $\sim65^{\circ}$.  
The jet emanates from the nucleus and can be traced out to kilo-pc-scales 
(Cohen et al. 1971; Wrobel 1984; Jones 1984, Kellermann et al. 1998;
Kameno et al. 2001; Vermeulen et al. 2003; Kadler et al. 2004b).
The lower limit to the jet inclination of 57$^{\circ}$ is estimated by Vermeulen et al. 2003. 
Radio observations of the low-luminosity active galactic nucleus (AGN) 
of NGC 1052 with VLBI at multiple frequencies have revealed the presence of a dense
circumnuclear structure, 
which obscures the very center of this elliptical galaxy 
(Kellermann et al. 1998; Kameno et al. 2001; Vermeulen et al. 2003;
Kadler et al. 2004b). 

This object shows a convex radio spectrum peaked at 10 GHz, to be classified as a GHz-Peaked Spectrum (GPS) source (de Vries et al. 1997; O'Dea 1998).  ~ 
The convex spectum is explained as free-free absorption (FFA) 
due to a dense plasma obscuring the nuclear torus 
(Kameno et al. 2001, 2003; Kadler et al. 2004b; Vermeulen et al. 2003).  
X-ray spectra also imply a high column density 
of $10^{22}$ cm$^{-2}$ to $10^{23}$ cm$^{-2}$
toward the center, and support the presence of a dense gas torus
(Guainazzi \& Antonelli 1999; Weaver et al. 1999; Kadler et al. 2004a).  

The center of NGC~1052 harbors a luminous H$_2$O megamaser, 
which is redshifted by 50--350 km~s$^{-1}$ with respect to 
the systemic velocity of the galaxy 
(1491~km~s$^{-1}$; de Vaucouleurs et al. 1991). 
The spectral profile typically shows a broad velocity 
width of $\sim$100~km~s$^{-1}$ (FWHM) (Braatz, Wilson $\&$ Henkel, 1994, 1996, Braatz et al. 2003).  ~ 
Past VLBI images reveal that 
H$_2$O maser gas with the velocity range of 1585--1685~km~s$^{-1}$ 
is distributed along the continuum ridge of the western jet, 
0.05--0.1~pc shifted to the west from the gap between the eastern and western jets
in November 1995. 
Excitation by shocks 
into the dense molecular clump which lies in or around the radio jet, 
or amplification of the radio continuum emission of the jet 
by foreground molecular clouds,  were suggested by Claussen et al. (1998). 
On the other hand, 
Kameno et al. (2005) presented the circumnuclear torus model to explain the time variability of 
the H$_2$O maser emission. 
Relevant to the interpretation of the H$_2$O maser emission line, 
several absorption lines are also found toward the center of NGC~1052 
(H~I: van Gorkom et al. 1986; OH : Omar et al. 2002; 
HCO$^{+}$, HCN and CO : Liszt and Lucas 2004). 

\section{Observations and Results}

In order to confirm the positional relation 
between the H$_2$O maser gas and the proposed circumnuclear torus, 
we observed the continuum and maser emissions in the nucleus 
of NGC~1052 with the VLBA. 
The continuum emissions were measured at 15 and 43 GHz. 
At 22 GHz, 
the H$_2$O maser emission and the line-free continuum emission were 
observed simultaneously. ~ 
Channel maps of H$_2$O maser emission were made every 6.74 km~s$^{-1}$, 
averaging every 10 spectral channels.

\begin{figure}
\includegraphics[scale=0.45]{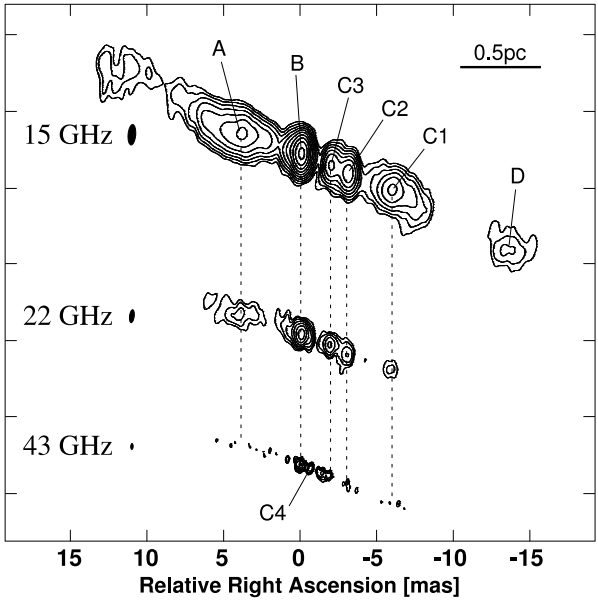}
\caption{The aligned images of NGC~1052 at 15, 22 and 43 GHz 
with the VLBA.  
The synthesized beam sizes (FWHM) are 1.3$\times$0.49 mas 
in P.A.$= -4.8^{\circ}$, 0.86$\times$0.32 mas in P.A.$=-7.1^{\circ}$, 
0.39$\times$0.17 mas in P.A.$=-2.8^{\circ}$, respectively, 
as shown in the left of each image.
The contours start at $3\sigma$ level, increasing by a factor of 2, 
where $\sigma=$ 0.24, 1.07 and 1.45~mJy~beam$^{-1}$, respectively 
at 15, 22 and 43~GHz. }
\label{fig:cont}
\end{figure}

The continuum images at 15, 22 and 43 GHz show the two-sided jet structure 
which consists of several components (Fig.~\ref{fig:cont}). 
The extended structures of knots A, C1 and D become fainter at higher frequency bands.
Knots B, C2 and C3 are resolved into several components at 22 and 43 GHz. 
The central engine is supposed to lie between the eastern jet (knots A and B) and 
the western jet (knots C3, C2, C1 and D). 
The 43~GHz image reveals another knot (C4) located between B and C3. 
The knot C4 is similar to the component A15 by Kadler et al. (2004b), 
which is supposed to be the innermost part of the eastern jet.
In the 43~GHz image, knot A has poorly defined morphology, 
and knot C1 is split into several faint peaks. 
These images at different frequencies include uncertainty 
in absolute positional information through the self-calibration process. 
For alignment of these images, we used the relative positions 
of knots B, C1 and C2, 
which are clearly seen in the restored images with a 1.30$\times$0.49 mas beam 
at 15, 22 and 43 GHz. 
Then we derived relative offsets 
to minimize the positional residuals 
(e.g. Kameno et al. 2001; Kameno et al. 2003; Sawada-Satoh et al. 2008). 
Finally, we could overlay the image at 15~GHz and 43~GHz with the image at 22~GHz,  
with positional errors of ($\pm$0.05, $\pm$0.04) mas and ($\pm$0.05, $\pm$0.09) mas, in R.A. and DEC respectively.

\section{Discussions}
\subsection{Free-Free Absorption Opacity Distribution}
After restoring with the same beam, we obtained spectral indices along the jets.
The spectral indices indicate 
that most parts of the two-sided jet structure have optically 
thin synchrotron spectra at 15--43~GHz 
except at the inner edge of the inner part of the both jets;  
a steeply rising spectrum 
($\alpha^{22}_{15}=3.2\pm0.1$, $\alpha^{43}_{22}=3.1\pm0.1$; $S\propto\nu^{\alpha}$)
at the western edge of knot B and 
at the eastern edge of knot C3 are revealed. 
The spectral index exceeds the theoretical limit for synchrotron self-absorption ($\alpha=2.5$). 
The highly rising spectrum of the inner edge of the both jets implies 
that the synchrotron emission is obscured 
through the free-free absorption (FFA) by the foreground dense plasma, 
and this is consistent with past multi-frequency observations 
(Kameno et al. 2001; Vermeulen et al. 2003; Kadler et al. 2004b).

\begin{figure}
\includegraphics[scale=0.45]{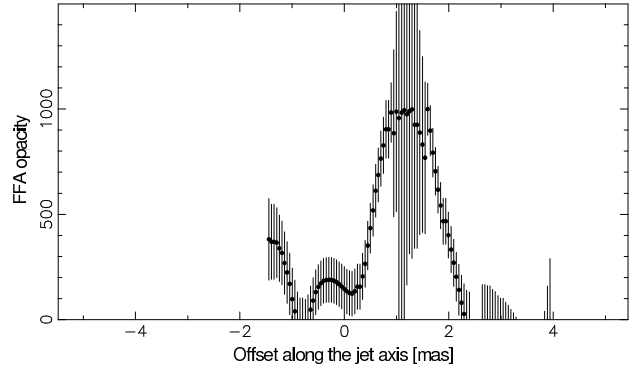}
\caption{Free-free absorption opacity along the jet axis (P.A. = 65$^{\circ}$)
which is shown as a solid line in Fig.~\ref{fig:contmas}. 
The fit in the inner edge has larger errors, 
because the continuum spectrum of the inner edge does not show 
the peak between 15 and 43 GHz. }
\label{fig:opacity}
\end{figure}

Fitting the continuum spectrum at 15--43~GHz to FFA model
($S_{\nu} \propto \nu^{- \alpha} \exp{(- \tau\nu^{-2.1})}$), 
we obtained FFA opacity ($\tau$) distributions along the jet axis (Fig.~\ref{fig:opacity}), 
which reveals that high opacity ($\tau > 1000$) is found in the inner edge of the jets.  
It implies that 
the dense cold plasma cover $\sim$2~mas in the inner edge, where the central engine is supposed to exist. 

\subsection{H$_2$O maser emission}

Significant H$_2$O maser emission within the velocity range of 1550--1850 km~s$^{-1}$ 
were detected, 50--350 km~s$^{-1}$ redshifted from the systemic velocity of the galaxy, and 
it is consistent with past single-dish observations 
(Braatz et al. 1994, 1996; Braatz et al. 2003; Kameno et al. 2005). 
The maser spots consist of two clusters; 
the eastern cluster and the western cluster are located 
on knots B and C3, respectively. 
The H$_2$O masers projected on the approaching jet, 
or the eastern cluster were detected for the first time. 
In the western cluster,  maser spots are distributed 
along the jet axis direction that span $\sim$1 mas (0.1 pc), 
and show some velocity shift along the jet axis direction. 
Position-velocity diagrams of the H$_2$O maser spots 
along the jet axis  (Fig.~\ref{fig:pv}b) also reveal the 
 trend of velocity gradient 
of $\sim$ 250 km~s$^{-1}$~mas$^{-1}$ 
in the western cluster. 
We note that Claussen et al. (1998) showed a velocity gradient 
along the east-west direction of magnitude $\sim$ 100 km~s$^{-1}$~mas$^{-1}$  
in the maser cluster on the western jet knot. 
A velocity gradient in the eastern cluster is not obvious (Fig.~\ref{fig:pv}a).  

\begin{figure}
\includegraphics[scale=0.38]{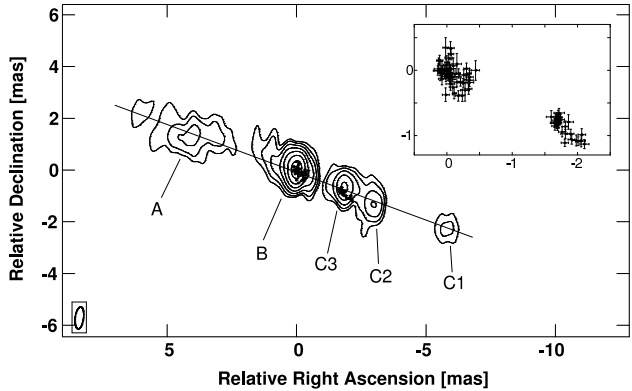}
\caption{Relative distributions of H$_2$O maser spots (full circle) 
  with respect to the continuum image at 22~GHz (contour). 
  The inset box shows the detailed distributions of maser spots.
  }
\label{fig:contmas}
\end{figure}

\subsection{Interpretation of the nuclear structure}

The brightness temperature ratio between the approaching and receding jets is  
related to the viewing angle of the jet axis and the true jet speed. 
Assuming that knots A and C1 form a symmetric pair of knots 
on either side of the nucleus, 
their intensity ratio $R$ is given by 
\begin{equation}
R=\frac{T_{b}^{A}}{T_{b}^{C1}}
= \biggl( \frac{1+\beta \cos{\theta}}{1-\beta \cos{\theta}} \biggr) ^{3-\alpha}
\end{equation}
where $T_{b}^{A}$ and $T_{b}^{C1}$ are the brightness temperature of knots A and C1, 
respectively, 
$\beta$ is the true jet velocity as a fraction of the speed of light ($v/c$), 
$\theta$ is the viewing angle of the jet axis, 
and $\alpha$ is the spectral index. 
Adopting  
$\beta=0.64$ (Kadler et al. 2004b) and $\alpha=-1$, 
 the viewing angle is estimated to be 79--80$^{\circ}$ and 76--90$^{\circ}$
 using the flux density at 15 and 22 GHz, respectively.
Kadler et al. (2004b) obtained the minimum value of $\beta=0.64$ 
by taking the jet inclination derived by Vermeulen et al. (2003). 
The jet axis is considered to be nearly parallel to the sky plane.

There are two main ideas about what the H$_2$O masers of NGC~1052 are 
associated with, the jet or the circumnuclear torus (Claussen et al. 1998; Kameno et al. 2005). 
Here we discuss the both ideas to account for the results from the observations. 
One possible scenario is that the H$_2$O masers are associated with the circumnuclear 
torus with the X-ray dissociation region (XDR). 
The H$_2$O masers are distributed  
where the FFA opacity is large, and 
this suggests that the H$_2$O masers and the plasma exist close to 
each other.  
In the case of the H$_2$O megamaser emission in NGC~4258, 
Neufeld \& Maloney (1995) and Herrnstein et al. (1996) 
proposed that the molecular disk consists of several layers including 
a heated molecular layer where the H$_2$O masers reside.  

Kameno et~al. (2005) applied this idea to the circumnuclear 
torus model in NGC 1052. 
A hot ($\sim$ 8000 K) plasma layer is created on the inner surface of the torus because of the direct exposure to the X-ray radiation from the 
central source. 
This layer is responsible for the free-free absorption. 
The XDR which lies immediately 
next to the plasma layer inside the torus, 
is heated above $\sim 400$~K as it is still partially irradiated 
by the X-ray radiation (Maloney 2002). 
Excited H$_2$O molecules 
in the XDR will amplify the continuum seed emission 
from the jet knots in the background and result in maser emissions. 
The presence of masers on both jets 
indicates the thickness of the torus along the orientation of the jet,   
covering at least knot B and C3. 
If the orientation of the jet axis is parallel to the sky plane, 
the thickness of the torus should be therefore 0.2~pc at least. 
More dominant FFA and H$_2$O masers on the receding jet 
support the circumnuclear torus scenario, 
since the path length within the torus toward the receding jet would be greater. 

\begin{figure}
\includegraphics[scale=0.35]{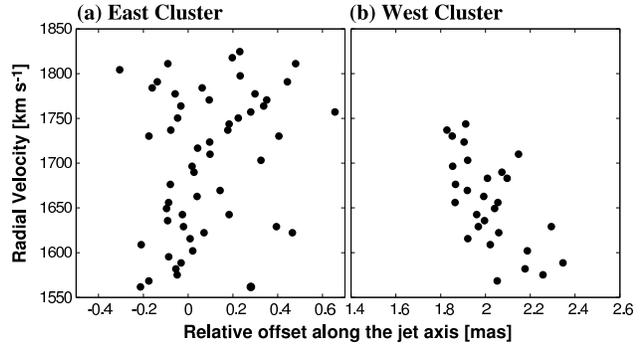}
\caption{Position-Velocity diagram of H$_2$O maser spots 
along the jet axis with respect to the peak of knot B.  
A clear velocity shift as a function of relative offset along the 
jet axis (P.A.=65$^{\circ}$ in Fig.~\ref{fig:opacity}) is seen in the western cluster (b). 
On the other hand, 
the eastern cluster seems to include several sub structures in velocity (a). 
}
\label{fig:pv}
\end{figure}

The FFA opacity $\tau$is a function of the electron density $n_e$ and temperature 
$T_e$ along the line of sight as 
\begin{equation}
\tau = 0.46 \int_{LOS} n^2_e T^{-1.5}_e dL
\end{equation}
The maximum opacity $\tau \sim$ 1000 in the path length of $\sim$ 1 pc
gives $n^2_e T^{-1.5}_e = 2100$.  
We can estimate the electron density $n_e \ge 4.6 \times 10^4$ cm$^{-3}$
using Ionization condition $T_e \ge 10^4$ K. 
The electron column density $n_e L$  would be 
$1.4 \times 10^{23} (T_e / 10^4 \rm{K})^{0.75}$ cm$^{-2}$, 
and the value is comparable to the atomic column density 
($\sim 10^{23}$ cm$^{-2}$) obtained by 
ROSAT and ASCA X-day observations (Guainazzi \&Antonelli 1999).  

If the H$_2$O masers are associated with the torus, 
the redshifted spectrum of the H$_2$O maser emission
would be accounted for by a contraction toward the central engine. 
The positional-velocity diagram along the jet axis 
for each maser cluster (figure~\ref{fig:pv}) 
gives the appearance 
that the H$_2$O maser gas closer to the central engine is more redshifted. 
Such a velocity shift as a function of positional offset 
could indicate the acceleration of the infalling gas 
toward the central engine. 
Thus, the circumnuclear torus scenario can explain the observed characteristics.

Alternatively, the H$_2$O masers could be excited 
by the interaction between the jet and circumnuclear molecular gas
as Claussen et al. (1998) proposed,  
and as interpreted in the case of Mrk~348 (Peck et al. 2003). 
A weak point of the jet excitation scenario is the difficulty to explain 
why the eastern cluster of H$_2$O masers projected 
against the approaching jet is also redshifted. 
For the masers in the western cluster moving with the most redshifted velocity 
($\sim$ 400 km~s$^{-1}$) from the systemic velocity of the galaxy, 
the maser gas should have moved 1.9$\times$10$^{-3}$ pc eastward 
from November 1995 to July 2000, 
because the jet axis is close to the sky plane.  
This motion is too small to detect for the five-year multi-epoch observations.  
Further VLBI observations are necessary 
in order to detect the proper motion of H$_2$O maser spots 
predicted by the jet excitation model, in order to support or 
refute this scenario.

\begin{figure}
\includegraphics[scale=0.39]{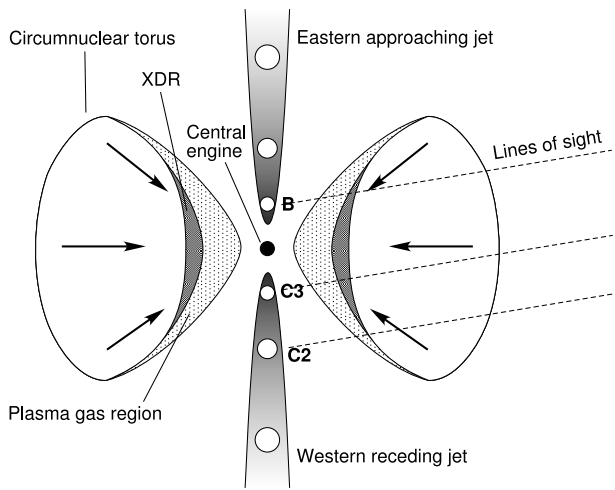}
\caption{A cartoon showing the possible environment in the 
circumnuclear torus and jets in NGC~1052. 
Double-sided jet axis inclined by $\ge 76^{\circ}$ 
 with respect to the line of sight. 
Inner surface of the torus is ionized by X-ray emission.
XDR is formed on the inner layer 
of the torus and amplifies background continuum emission 
from the jet knot. 
On B and C3, we can see the H$_2$O maser emission and FFA absorption as well. 
On the other hand, only FFA appears on C2. 
Since the gas inside the torus is falling toward the central 
engine,  redshifted H$_2$O maser emission is detected.  
}
\label{fig:torusmodel}
\end{figure}

\acknowledgements
The VLBA is operated by the NRAO, a facility of the National Science Foundation operated under cooperative agreement by Associated Universities, Inc. 
\newpage


\begin{thebibliography}{}
\bibitem{} Braatz, J.A., Wilson, A. S., Henkel, C.: 1994, \apj, 437, L99.
\bibitem{} Braatz, J.A., Wilson, A. S., Henkel, C.:1996, \apjs, 106, 51.
\bibitem{} Braatz, J.A., Wilson, A. S., Henkel, C., Gough, R., Sinclair, M.:2003, \apjs, 146, 249.
\bibitem{} Claussen, M., Diamond, P. J., Braatz, J. A., Wilson, A. S., Henkel, C.: 1998, \apj, 500, L129.  
\bibitem{}Cohen, M. H., Cannon, W., Purcell, G. H., Shaffer, D. B., Broderick, J. J., Kellermann, K. I.,  Jauncey, D. L.: 1971, \apj, 170, 207.
\bibitem{} de Vaucouleurs, G. et al.: 1991, 
Third Reference Catalogue of bright galaxies.
\bibitem{} de Vries, W. H., Barthel, P. D., O'Dea, C. P.: 1997, A$\&$A, 321,105
\bibitem{} Gabel, J. R., Bruhweiler, F. C., Crenshaw, D. M., Kraemer, S. B., Miskey, C. L.: 2000, \apj, 532, 883.
\bibitem{} Guainazzi, M.,  Antonelli, L. A.: 1999, \mnras, 304, L15.
\bibitem{} Herrnstein, J. R., Greenhill, L. J., Moran, J. M.: 1996, \apj, 468, L17. 
\bibitem{} Jones, D., 1984, \apj 276, 480.
\bibitem{}  Kadler, M., Kerp, J., Ros, E., Falcke, H., Pogge, R.W., Zensus, J. A.: 2004a, A$\&$A, 420, 467
\bibitem{}  Kadler, M., Ros, E., Lobanov, A. P., Falcke, H., Zensus, J. A. 2004b, A$\&$A, 426, 481
\bibitem{} Kameno, S., Sawada-Satoh, S.,
Inoue, M., Shen, Z.-Q., Wajima, K.: 2001,\pasj, 53, 169. 
\bibitem{} Kameno, S., Inoue, M., Wajima, K., Sawada-Satoh, S., 
Shen, Z.-Q. : 2003, Publ. Astron. Soc. Australia, 20, 134.
\bibitem{} Kameno, S, Nakai, N.,Sawada-Satoh, S., Sato, N., Haba, A.: 2005,\apj, 620,  145.
\bibitem{} Kellermann, K. I., Vermeulen, R. C., Zensus, J. A., Cohen, M. H.:1998, \aj, 115, 1295.
\bibitem{}Knapp G.R. Faber, S.M., Gallagher, J.S: 1978, \aj, 83, 139 
\bibitem{} Liszt, H., Lucas, R.: 2004, A$\&$A, 428, 445.
\bibitem{} Malony, P.R.: 2002, PASP,19,401.
\bibitem{} Neufeld, D.A., Malony, P.R.: 1995, \apj, 447, L17. 
\bibitem{} O'Dea, C.P.: 1998, PASP, 110, 493.
\bibitem{}  Omar, A., Anantharamaiah, K. R., Rupen, M., Rigby, J.: 2002, A$\&$A, 381, L29.
\bibitem{}Peck, A. B., Henkel, C., Ulvestad, J. S., Brunthaler, A., Falcke, H., Elitzur, M., Menten, K. M., Gallimore, J. F.: 2003, \apj, 
590, 149.
\bibitem{} Sawada-Satoh, S.,  Kameno, S., Nakamura, K., Namikawa, D., Shibata, K.M.; Inoue, M.: 2008, \apj, 680, 191.
\bibitem{} Spergel, D.N., Verde, L., Peiris, H. V. et al.: 2003, \apjs, 148, 175. 
\bibitem{}  van Gorkom, J. H., Knapp, G. R., Raimond, E., Faber, S. M., Gallagher, J.S.:  1986, \aj, 91, 791.
\bibitem{} Vermeulen, R. C., Ros, E., Kellermann, K. I., Cohen, M. H., Zensus, J. A., van Langevelde, H. J.: 2003, A$\&$A, 401, 113.
\bibitem{}Weaver, K. A., Wilson, A. S., Henkel, C.,  Braatz, J. A.: 1999, \apj, 520, 130
\bibitem{} Wrobel, J. M.: 1984, \apj, 284, 531.
\end{thebibliography}
\end{document}